\documentclass[useAMS,usenatbib,fleqn]{mn2e}

\usepackage{graphicx,color}
\usepackage{amsmath}
\usepackage{amssymb}
\usepackage{bm}
\usepackage{ulem}
\usepackage{enumerate}
\usepackage{extarrows}
\usepackage{times}

\renewcommand{\vec}[1]{{{\mbox{\boldmath $#1$}}}}

\newcommand{\aar}{\mathrm{,a}}

\newcommand{\iir}{\mathrm{,i}}

\newcommand{\Exp}[1]{{\rm e}^{#1}}

\newcommand{\del}{\partial}
\newcommand{\Del}{{\nabla}}

\newcommand{\bmDel}{\bm{\nabla}}

\newcommand{\eps}{\epsilon}

\newcommand{\Emf}{\bm{\mathcal{E}}}

\newcommand{\Flux}{\bm{\mathcal{F}}}

\newcommand{\bfu}{\bm{u}}

\newcommand{\bmb}{\bm{b}}

\newcommand{\mean}[1]{\overline{#1}}

\newcommand{\corot}{_\mathrm{c}}						
\newcommand{\D}{_\mathrm{D}}						
\newcommand{\eq}{_\mathrm{eq}}						
\newcommand{\f}{_\mathrm{0}}					   	
\newcommand{\kin}{_\mathrm{k}}			   		
\newcommand{\magn}{_\mathrm{m}}			   		
\newcommand{\crit}{_\mathrm{c}}			   		
\newcommand{\ma}{_\mathrm{max}}			   		

\newcommand{\on}{_0}

\newcommand{\cro}{\times}

\renewcommand{\theenumi}{\roman{enumi}}

\newcommand{\pat}{_\mathrm{p}}
\newcommand{\arm}{_\mathrm{a}}
\newcommand{\interarm}{_\mathrm{i}}
\newcommand{\critarm}{_\mathrm{c,a}}
\newcommand{\critinterarm}{_\mathrm{c,i}}
\newcommand{\uarm}{_\mathrm{U,a}}
\newcommand{\uinterarm}{_\mathrm{U,i}}
\newcommand{\kappaarm}{_\mathrm{\kappa,a}}
\newcommand{\kappainterarm}{_\mathrm{\kappa,i}}

  \newcommand{\kms}{\,{\rm km\,s^{-1}}}
  \newcommand{\kmskpc}{\,{\rm km\,s^{-1}\,kpc^{-1}}}

  \newcommand{\kpc}{\,{\rm kpc}}

  \newcommand{\Myr}{\,{\rm Myr}}
  
  \newcommand{\Gyr}{\,{\rm Gyr}}

\definecolor{webgreen}{rgb}{0,.5,0}
\definecolor{webbrown}{rgb}{.6,0,0}
\definecolor{purple}{rgb}{0.5,0,.5}

\title{Magnetic spiral arms and galactic outflows}
\author[L.\ Chamandy, A.\ Shukurov \& K.\ Subramanian]{Luke Chamandy$^{1}$\thanks{E-mail: luke@iucaa.ernet.in;
anvar.shukurov@ncl.ac.uk; kandu@iucaa.ernet.in}, 
Anvar Shukurov$^{2,1},$ \& Kandaswamy Subramanian$^{1}$\\
$^{1}$Inter-University Centre for Astronomy and Astrophysics, Post Bag 4, Ganeshkhind, Pune 411007, India\\
$^{2}$School of Mathematics \& Statistics, Newcastle University, Newcastle upon Tyne NE1 7RU}

\begin{document}


\pagerange{\pageref{firstpage}--\pageref{lastpage}} \pubyear{2014}

\maketitle

\label{firstpage}

\begin{abstract}
Galactic magnetic arms have been observed between the gaseous arms of some spiral galaxies; 
their origin remains unclear. 
We suggest that magnetic spiral arms can be naturally generated in the interarm regions 
because the galactic fountain flow or wind is likely to be weaker there than in the arms.
Galactic outflows lead to two countervailing effects:
removal of small-scale magnetic helicity, which helps to avert catastrophic quenching of the dynamo,
and advection of the large-scale magnetic field, which suppresses dynamo action.
For realistic galactic parameters, the net consequence of outflows being stronger in the gaseous arms 
is higher saturation large-scale field strengths in the interarm regions as compared to in the arms.
By incorporating rather realistic models of spiral structure and evolution into our dynamo models,
an interlaced pattern of magnetic and gaseous arms can be produced.
\end{abstract}
\begin{keywords}
magnetic fields -- dynamo -- galaxies: magnetic fields -- galaxies: spiral
\end{keywords}

\section{Introduction}
Magnetic arms are spiral-shaped segments of enhanced large-scale magnetic field, 
sometimes observed between the optical (and gaseous) arms of spiral galaxies. 
They were first observed in the galaxy IC~342 \citep{Krause93} 
and later 
in the galaxy NGC~6946 \citep{Beck+Hoernes96}. 
In NGC~6946, magnetic arms have pitch angles similar to those of the gaseous arms; 
they lag the gaseous arms by 20--$50^\circ$, 
so that the two spiral systems apparently do not intersect \citep{Frick+00}. 
In M51, they partially overlap with the optical arms 
showing an intertwined pattern \citep{Fletcher+11}. 
This behaviour is opposite to what is expected of a frozen-in magnetic field, 
whose strength increases with gas density, 
and thus favours a dynamo origin of the large-scale galactic magnetic fields. 
It is not quite clear how widespread is this phenomenon among spiral galaxies and what causes it.

Several attempts have been made to explain interarm large-scale fields
in the context of the mean-field dynamo theory. 
\citet{Lou+Fan98} addressed this problem with a model of MHD spiral density waves, 
but using rather simplified configurations of magnetic fields and galactic rotation curves.
\citet{Moss98} took the turbulent magnetic diffusivity to be larger in the gaseous arms 
or the $\alpha$ effect to be weaker. 
\citet{Rohde+99} assumed that the correlation time of interstellar turbulence is larger within the gaseous arms.
However, weak, if any, 
observational or theoretical evidence is available to support the assumptions used in these models. 

From the generic form of the dynamo non-linearity, \citet{Shukurov98} concluded
that the steady-state large-scale magnetic field can be stronger between the
gaseous arms if the dynamo number is close to its critical value for the 
dynamo action, i.e., magnetic arms interlaced with gaseous arms
can occur in galaxies with a weak dynamo action. 

\citet{Moss+13} argued that stronger turbulence in the gaseous arms, 
driven by higher star formation rate, 
would produce stronger turbulent magnetic fields leading to the saturation 
of the large-scale dynamo at a lower level.  
However, stronger star formation may not lead to stronger turbulent motions in the warm phase, 
the site of the large-scale dynamo action. 
The turbulent velocity is limited by the sound speed in the warm gas 
($\sim10\kms$ for a wide range of star formation rates). 
Extra energy injected by enhanced star formation mostly feeds the hot phase 
of the interstellar gas and enhances gas outflow from the galactic disc. 
Moreover, the effects of the small-scale field on the dynamo 
could be more subtle \citep{Brandenburg+Subramanian05a} than envisaged by \citet{Moss+13}.
  
\citet{Chamandy+13a,Chamandy+13b} extended the mean-field dynamo equations
to allow for a finite relaxation time $\tau$ of the turbulent electromotive force. 
This results in dynamo equations that admit wave-like solutions and thus might produce magnetic arms 
in a way reminiscent of how a spiral pattern is produced by density waves.
However, the wave-like behaviour turns out to be insignificant for realistic values of $\tau$. 
If the $\alpha$ effect is stronger in the gaseous arms, 
a finite $\tau$ of a realistic magnitude can produce magnetic arms lagging by 
$(30$--$40)^\circ$ behind the respective gaseous arms, 
similar to what is observed in NGC~6946. 

Most of the above models use a rigidly rotating spiral,
and thus only produce non-axisymmetric fields near the corotation radius,
because of strong differential rotation of the gas.
This is at odds with observations showing radially extended magnetic arms in some galaxies.
Such a feature can be reconciled with more modern theories of spiral structure, 
where the spirals can wind up or can be due to interfering rigidly rotating patterns \citep{Chamandy+13a,Chamandy+14a}.
\citet{Kulpadybel+11} observed a systematic drift of magnetic arms from the gaseous arms in their numerical model
with an evolving spiral pattern.
These results suggest that magnetic arms should be studied in a broader context of diverse spiral pattern models.

A simple, natural and direct effect of the gaseous spiral arms on the mean-field dynamo
was suggested by \citet{Sur+07} who found that dynamo action is sensitive 
to galactic outflows.
This links the dynamo efficiency directly to star formation rate which is confidently known to be higher 
within the gaseous arms.
In the Milky Way, OB associations that drive gas outflows concentrate in spiral arms \citep{Higdon+Lingenfelter13}. 
The H\,{\sc i} holes caused by hot superbubbles and chimneys, and the vertical flows driven by them, 
tend to be concentrated in the spiral arms of NGC~6946 \citep{Boomsma+08}. 
There is also evidence for non-axisymmetric distributions of the extra-planar \mbox{H\,{\sc i}} in other galaxies, 
suggesting a non-axisymmetric outflow pattern \citep{Kamphuis+13}.
The purpose of this Letter is to demonstrate, using a nonlinear galactic dynamo model
incorporating realistic spiral evolution models,
that magnetic arms can be sustained between the gaseous arms due to a stronger gas outflow along the gaseous spiral. 

\section{Galactic dynamo with an outflow}\label{sec:outflow}
Galactic outflows (winds or fountains) facilitate the mean-field dynamo action by 
removing helical turbulent magnetic fields from the disc 
which would have otherwise catastrophically quenched the dynamo \citep{Shukurov+06}. 
However, as outflows also remove the total field,
the dynamo is damaged if the outflow is too strong,
so there is an optimal range of outflow speeds.
In addition, other fluxes of magnetic helicity, 
such as a turbulent diffusive flux, may also help to avert catastrophic quenching.
These effects are captured by a simple, yet remarkably accurate approximate
solution of the dynamo equations, the no-$z$ approximation \citep{Chamandy+14b}. 
This solution yields the following expression for the steady-state (saturated) strength of the mean magnetic field 
$\vec{B}$ written in the cylindrical frame $(r,\phi,z)$ with the origin 
at the galactic centre and the $z$-axis aligned with the angular velocity $\vec{\Omega}$
\citep{Chamandy+14b}:
\begin{equation}
  \label{Bsteady}
  B^2\simeq K(R_U+\pi^2 R_\kappa)\left({D}/{D\crit}-1\right),
\end{equation}
with
\begin{equation}
  \label{K}
  K={B\eq^2}l^2/[2h^2\xi(p_B)]\,.
\end{equation}
Here $R_U=U_zh/\eta$ is a dimensionless measure of the outflow intensity,
with $U_z$ the mass-weighted mean vertical velocity
and $h$ the scale height of the dynamo-active layer. 
Further, $\eta\simeq\tfrac{1}{3}lu$ is the mean-field turbulent diffusivity, 
with $l$ the turbulent scale and $u$ the rms turbulent speed.
$R_\kappa\equiv\kappa/\eta$ is the ratio of the 
diffusivity
of the mean current helicity 
and 
the diffusivity of the magnetic field. 
Denoting the gas density by $\rho$,
$B\eq=(4\pi\rho u^2)^{1/2}$
is the field strength for which there is energy equipartition between magnetic field and turbulence.
For convenience we have also made use of the notation
$\xi(p_B)\equiv1-3\cos^2 p_B/(4\sqrt2)$,
where $p_B$ is the magnetic pitch angle,
and is related to the magnetic field by the expression $\tan p_B=B_r/B_\phi$.
Finally, $D\approx 9(\Omega h^2/u^2)d\Omega/d\ln r<0$ is the dynamo number \citep{Ruzmaikin+88},
a dimensionless measure of the induction effects of differential rotation and helical turbulence.
There is a critical dynamo number $D\crit$, such that the kinematic growth rate of the mean magnetic field is positive (negative)
for $D/D\crit>1$ ($<1$). 
Thus, equation~\eqref{Bsteady} applies for $D/D\crit\ge1$; otherwise $B$ is effectively zero.
In the no-$z$ approximation, the critical dynamo number is given by
\begin{equation}
  \label{Dcrit}
  D\crit\simeq-(\pi/2)^5(1+R_U/\pi^2)^2.
\end{equation} 

\begin{figure}                     
  \begin{center}
    \includegraphics[width=\columnwidth]{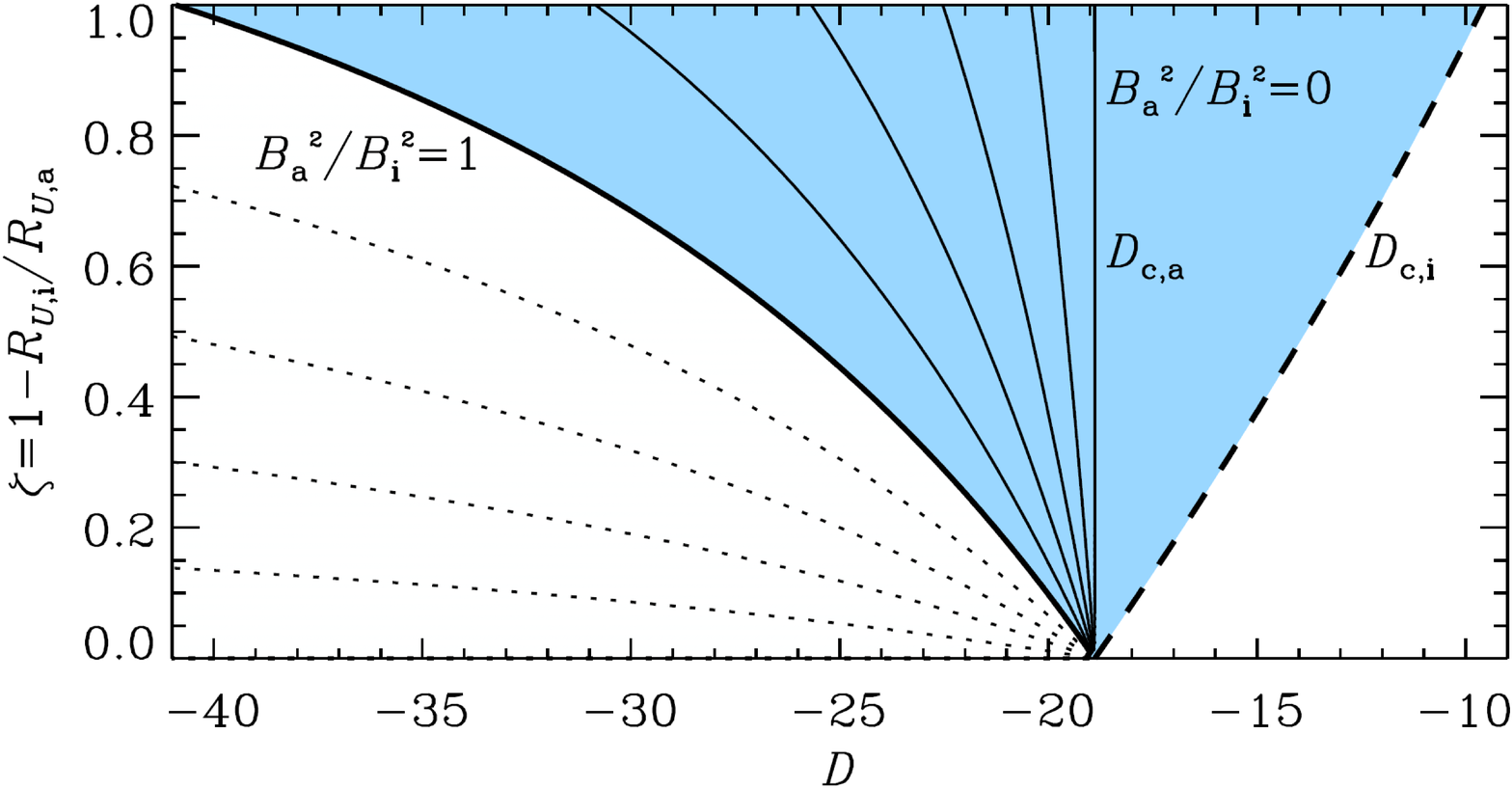}
    \caption{\label{fig:Bai}
      Contours of the arm-interarm contrast in magnetic field strength, $B\arm^2/B\interarm^2$,
      in the $(D,\zeta)$-plane for $K\arm/K\interarm=2$, $R_{\mathrm{U,a}}=4$ and $R_{\kappa\aar}=R_{\kappa\iir}=1$. 
      Contours are drawn with a spacing of $0.2$;
      those with $B\arm^2/B\interarm^2>1$ are shown as dotted lines.
      The dashed line shows $D\critinterarm$.
      In the shaded region, the dynamo is supercritical in the interarm regions,
      and the large-scale magnetic field is stronger in the interarm regions as compared to in the arms.
    }            
  \end{center}
\end{figure}

The quantity $R_U$ is expected to be closer to its optimum value for dynamo action in between the gaseous arms than within them,
causing the large-scale magnetic field to concentrate in the interarm regions.
To verify and test this idea, consider whether this can occur for realistic values of galactic parameters. 
We denote with subscripts `a' and `i' quantities within the gaseous arms and between them 
and for convenience we define the parameter
$\zeta\equiv 1-R\uinterarm/R\uarm$.
$\zeta$ is a measure of the arm--interarm contrast in the outflow speed;
it vanishes if there is no such contrast, 
and 
$\zeta=1$ if there is no outflow in the interarm regions.
It is reasonable to take $\eta\interarm=\eta\arm$
and $h\interarm=h\arm$; 
from equation~\eqref{K} for $K$ we then have 
$K\arm/K\interarm=(\xi\interarm/\xi\arm)(\rho\arm/\rho\interarm)$.
The first of these ratios $\simeq1$, which leaves $\rho\arm/\rho\interarm>1$.
We adopt 
$K\arm/K\interarm=2$ in the illustrative example 
of this section,
and we also set $D\arm=D\interarm=D$.
We then obtain from equation~\eqref{Bsteady} the arm--interarm contrast in the magnetic field strength:
\begin{equation*}
  \label{contrast}
  \frac{B\arm^2}{B\interarm^2}= 
                                \frac{K\arm}{K\interarm}\left[\frac{R\uarm+\pi^2 R\kappaarm}{(1-\zeta)R\uarm+\pi^2R\kappainterarm}\right]
                                \left(\frac{D/D\critarm-1}{D/D\critinterarm-1}\right).
\end{equation*}
Figure~\ref{fig:Bai} shows the contours of $B\arm^2/B\interarm^2$ in the $(D,\zeta)$-plane
for $K\arm/K\interarm=2$, $R\uarm=4$ and $R\kappainterarm=R\kappaarm=1$.
A thick solid line shows where $B\arm^2/B\interarm^2=1$,
while the $B\arm^2/B\interarm^2=0$ contour is located at $D=D\critarm\simeq-18.9$.
Contours for $B\arm^2/B\interarm^2>1$ are shown dotted, and $B\arm^2/B\interarm^2=2$ traces the ($D<D\critarm$ part of the) $D$ axis.
The interarm critical dynamo number $D\critinterarm$ is shown by a dashed line.
The dynamo is supercritical in the interarm regions ($D/D\critinterarm>1$)
and the interarm saturated field exceeds that in the arms ($B\arm^2/B\interarm^2<1$)
for the shaded region of the parameter space of Fig.~\ref{fig:Bai}.

Note that concentration of magnetic field in the interarm regions becomes more likely with increasing $\zeta$.
Raising $R\uarm$ causes $|D\critarm|$ to increase, shifting the contours to the left
and enlarging the region of $(D,\zeta)$ space that satisfies the above conditions
($D\critinterarm|_{\zeta=1}=-(\pi/2)^5\simeq-9.6$ is not affected).
On the other hand, the effect of changing the ratio $K\arm/K\interarm$ is just to relabel the contours so,
e.g., halving $K\arm/K\interarm$ causes the contour $B\arm^2/B\interarm^2=1$ to become $B\arm^2/B\interarm^2=0.5$,
making the condition $B\arm^2/B\interarm^2<1$ easier to satisfy.

We see then that a larger value of the magnetic field in the interarm regions 
compared to that in the arms is possible for realistic dynamo parameters.
Note that the mechanism is most effective for dynamo numbers close to critical, large ratios of the arm/interarm outflow speeds,
and large (in absolute terms) outflow speeds in the arms.
Thus, magnetic arms can be displaced from the gaseous ones in galaxies with a relatively weak large-scale dynamo action. 
We now put these ideas on a firmer footing by considering a more detailed numerical model.

\section{Global galactic dynamo model}\label{sec:model}
With the velocity and magnetic fields split into the mean and fluctuating parts, 
$\bm{U}+\bm{u}$ and $\bm{B}+\bm{b}$ respectively, the induction equation 
averages to \citep{Moffatt78}
\begin{equation}
  \label{meaninduction}
    \del_t \vec{B} = \bmDel \times \left( \vec{U} \times \vec{B} + {\Emf}\right)\,,
  \quad
\Emf=\mean{\bfu\cro\bmb}\,,
\end{equation}
with $\del_t\equiv\del/\del t$ and $\Emf$ the mean electromotive force 
(bar denoting ensemble averaging) that solves \citep{Blackman+Field02}
\begin{equation}
  \label{minimaltau}
  \del_t \Emf=\tau^{-1}(\alpha\vec{B}-\eta \bmDel\times\vec{B}-\Emf)\,,
\end{equation}
where $\tau$ is the response time of $\Emf$ to changes in $\vec{B}$, $\vec{u}$ and $\vec{b}$.
The $\alpha$ effect is written as the sum of kinetic and magnetic contributions, 
$\alpha=\alpha\kin+\alpha\magn$, with $\alpha\kin\simeq l^2\Omega/h$ \citep{Krause+Radler80},
and \citep{Kleeorin+00,Subramanian+Brandenburg06,Shukurov+06}
\begin{equation}
  \label{dalpha_mdt}
  \del_t\alpha\magn
  =-2\eta\Emf\cdot\vec{B}/(l^2B\eq^2)-\bmDel\cdot\Flux\,,
\end{equation}
where $\Flux$ is the flux density of $\alpha\magn$ and the remaining notation is introduced in Section~\ref{sec:outflow}.
For $\tau\rightarrow0$, equation \eqref{minimaltau} reduces to the standard expression
$\Emf= \alpha\vec{B} -\eta\Del\cro\vec{B}.$

The advective and diffusive helicity transport give 
$\Flux= (\vec{U} -\kappa\bmDel)\alpha\magn$; it is reasonable to expect 
$\kappa=\mathcal{O}(1)\eta$. Limited numerical experiments suggest $\kappa=0.3\eta$ 
\citep{Mitra+10}, and $\kappa=0$ is included to explore the parameter space.

We solve equations \eqref{meaninduction}--\eqref{dalpha_mdt}
numerically using the thin disc approximation \citep{Ruzmaikin+88} and
the no-$z$ approximation \citep{Subramanian+Mestel93}, 
proved to be adequate in galactic discs \citep{Chamandy+14b},
which approximates the derivatives of $\vec{B}$ in $z$ by suitable ratios
of $\vec{B}$ to $h$, but retains the derivatives in $r$ and $\phi$.
The equations are solved on a polar grid of $200\times180$ mesh points in $r\times\phi$,
with $B_r=B_\phi=\del\alpha\magn/\del r=0$ at $r=0$ and $r=R$, and  
$B_r=-B_\phi=0.05 B\eq r(1-r^2)\Exp{-r/R}$ and $\alpha\magn=0$ at $t=0$; 
the results are not sensitive to the specific form of the initial conditions.
$|B_z|$ can be estimated from the condition $\bmDel\cdot\vec{B}=0$,
and turns out to be negligible for $r>1\kpc$, where the thin disc approximation is valid.

We use $l=0.1\kpc$, $u=10\kms$ \citep{Beck+96} 
and the Brandt rotation curve, $\Omega=\Omega\f/\sqrt{1+r^2/r_\omega^2}$;
$\Omega\f\simeq127\kms\kpc^{-1}$ and $r_\omega=2\kpc$ yield $U_\phi=250\kms$ 
at $r=10\kpc$. The ionised disc is assumed to be flared, similarly to the H{\sc i} layer,
$h=h\D\exp[(r-8\kpc)/r\D]$ with 
$h\D=0.5\kpc$ and $r\D=10\kpc$ \citep{Kalberla+Dedes08,Westfall+11,Eigenbrot+Bershady13,Hill+14},
but we also consider models with a flat ionised layer to confirm that this affects our
results insignificantly.
The equipartition field is assumed to vary with $r$ as $B\eq=B\f\exp(-r/R)$, $R=15\kpc$ \citep{Beck07}. 
The azimuthally averaged mean vertical velocity $U\f=3\kms$ is taken to be independent of $r$,
consistently with $U_z\approx0.2$--$2\kms$ of \citet{Shukurov+06}.
The spiral modulation of $U_z$, which is the key ingredient in the model, 
is discussed in Section~\ref{sec:spiral}.

\subsection{Models of the galactic spiral}
\label{sec:spiral}
Two models of spiral structure and evolution are explored;
they are chosen so as to be broadly consistent 
with the modern understanding of galactic spirals \citep[][and references therein]{Dobbs+Baba14}.
Both spiral models have trailing gaseous arms implemented via
an enhanced mean vertical velocity $U_z=U\f\widetilde{U}(r,\phi)$ in the arms,
where $U\f$ is the velocity amplitude 
and $\widetilde{U}(r,\phi)$ prescribes its spatial variation. 
For simplicity,
parameters other than $U_z$ do not vary with $\phi$ in these models.
Model~I, has two superposed
logarithmic spiral patterns that rotate rigidly at distinct angular velocities 
$\Omega_{\mathrm{p,1}}$ and $\Omega_{\mathrm{p,2}}$.
Here $\widetilde{U}= \max[1 +\epsilon\cos\chi_1 +\epsilon\cos\chi_2,\;0]$,
with $\chi_i=n_i(\phi -\Omega_{\mathrm{p,i}} t) -k_i\ln(r/R)$.
We generally take $\epsilon=1$ but try other values as well.
The inner spiral has two arms, $n_1=2$, with the corotation radius 
$r_{\mathrm{cor,1}}=6\kpc$ (giving $\Omega_{\mathrm{p,1}}\simeq40\kmskpc$),
and the outer one is three-armed, $n_2=3$, with $r_{\mathrm{cor,2}}=7\kpc$
($\Omega_{\mathrm{p,2}}\simeq35\kmskpc$).
We choose $k_1=-3$, producing the pitch angle $p=\tan^{-1}(n/k)\simeq-34^\circ$
of the inner spiral, and $k_2=-6$, so that $p\simeq-27^\circ$ in the outer region,
with $p<0$ corresponding to a trailing spiral.
A similar model is used in \citet{Chamandy+14a}; 
motivation for the parameter values adopted can be found there and in the references therein.

Model~D has an evolving two-armed pattern with a variable pitch angle 
\footnote{See equation (6.78) of 
\citet{Binney+Tremaine08}.
The negative sign on the right hand side is included here because we define trailing 
spirals to have $p<0$.}
$d(\cot p)/dt= -\Omega\pat$ and $\widetilde{U}= 1 +\epsilon\cos\lambda$, 
where $\lambda=2\{\phi -\Omega\pat [1 -\ln(r/r\corot)] (t -t\on)\}$,
$\Omega\pat\simeq31\kmskpc$ (so $r\corot=8\kpc$), and $\epsilon=1$.
At $t<t\on$, the disk is axisymmetric and the dynamo is already in a steady state;
the spiral pattern is turned on from a `bar' configuration at $t\on=10\Gyr$. 
(Winding up spiral arms are indeed found in simulations.)
The spiral's amplitude is modulated by
$\exp\left\{-[\log(r/r\f)]^2/(2w^2)\right\}$ with $r\f=6.5\kpc$ and $w=0.2$.
We have explored variations on this model to allow for a range of effects:
(i)~the amplitude truncated at $r\geq r\corot$
to account for the `forbidden' region around the corotation radius,
(ii)~$r\f$ increasing with time
at a speed $10$--$20\kms$ to approximate a travelling wave packet,
and (iii)~varying the amplitude in time
as in the swing amplification mechanism.
Since none of the modifications had a large impact on the magnetic pattern,
we only present results from the simpler model.

\begin{figure}    
\centering  
  \mbox{}\hspace*{-4.5em}
  \includegraphics[width=50mm,clip=true,trim= 00 0 0 5.5]{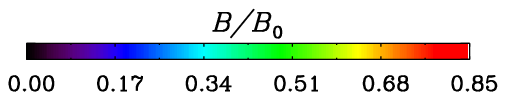}\\             
  \mbox{}\hspace*{-4.5em}
  \includegraphics[width=50mm,clip=true,trim= 0  0 0 13  ]{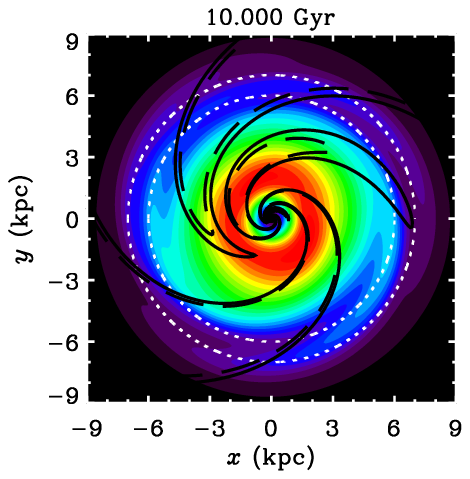}\\
  \mbox{}\hspace*{ 4.5em}
  \includegraphics[width=50mm,clip=true,trim= 00 0 0 5.5]{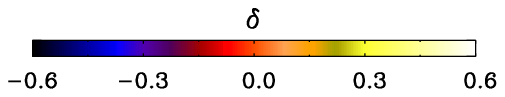}\\
  \mbox{}\hspace*{ 4.5em}
  \includegraphics[width=50mm,clip=true,trim= 0  0 0 13  ]{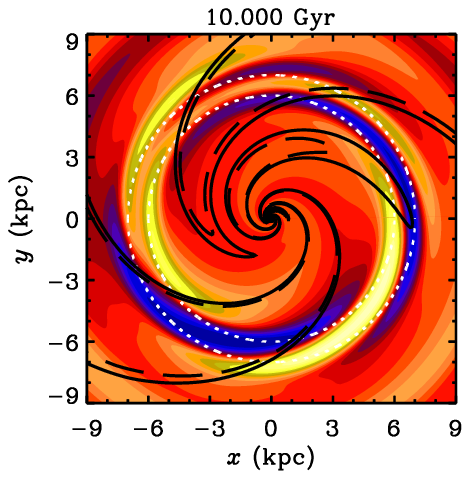}
  \caption{Large-scale magnetic field at $t=10\Gyr$ in Model~I, an interfering spiral pattern model.
Top: relative magnetic field strength $B/B_0$, where $B_0=B\eq|_{r=0}$. 
The corotation circles for the inner and outer patterns are shown dotted white,
the gaseous spirals are shown with black contours at $U_z/\max_\phi(U_z)=0.5$ (dashed)
and $0.4$ (solid). Bottom: the degree of non-axisymmetry $\delta=(B_\phi-B_\phi^{(0)})/B_\phi^{(0)}$.
\label{fig:I}
          }            
\end{figure}                       
\begin{figure}  
\centering                   
  \includegraphics[width=50mm,clip=true,trim= 0  0 0 13  ]{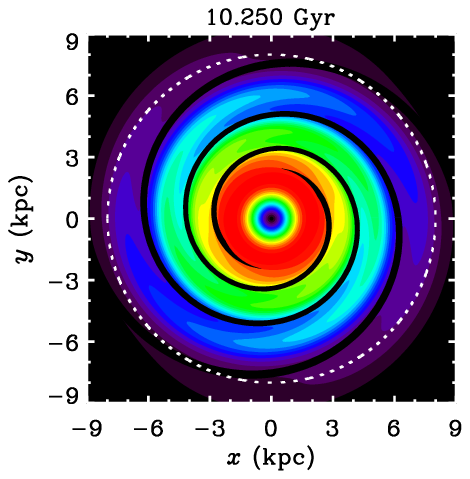}\\
  \includegraphics[width=50mm,clip=true,trim= 0  0 0 13  ]{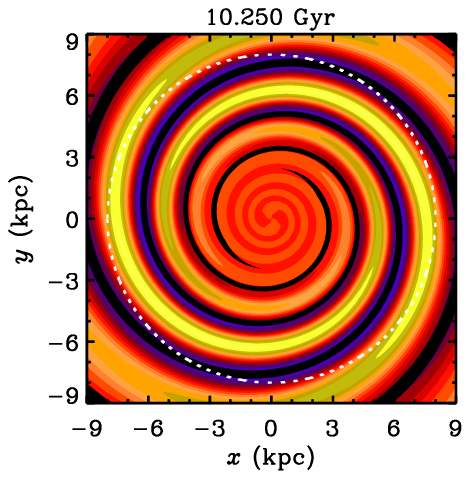}
  \caption{As in Fig.~\ref{fig:I} but for the linear spiral density wave model, Model~D, at $t-t\f=250\Myr$. 
                      The peak of the gaseous spiral ($U_z$) is shown by a filled 
                      black contour at $0.96\max(U_z)$ wherever it exceeds $0.1U\f$;
                      the colour scales are as in Fig.~\ref{fig:I}.
\label{fig:D}
          }            
\end{figure}                       
\begin{figure}
\centering                     
  \includegraphics[width=0.8\columnwidth,clip=true,trim= 0  0 0 0  ]{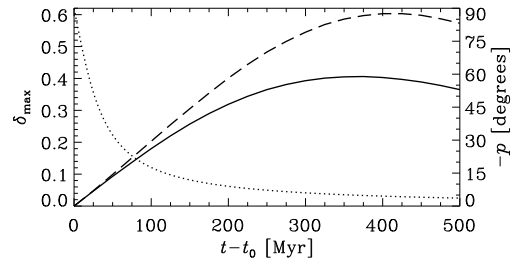}
  \caption{The evolution of the maximum degree of deviations from axial symmetry $\delta\ma$ in Model~D, 
           for $\tau\rightarrow0$ (solid) and $\tau=l/u\simeq10\Myr$ (dashed).
           The magnitude of the pitch angle of the gaseous spiral (right-hand axis) is shown dotted.
           \label{fig:delta_time}
          }            
\end{figure}                       

\section{Results}\label{sec:results}

Steady-state solutions presented here are obtained at $t\approx10\Gyr$
for $U\f=3\kms$, $\eps=1$, $\kappa=\eta$, and $\tau=0$ unless stated otherwise.
The top panel of Fig.~\ref{fig:I} shows magnetic field strength in Model~I normalised
to the equipartition value $B\eq$ at $r=0$.
The field strength is close to $B\eq$ near the centre (red--yellow colour) and smaller at larger $r$.
The magnetic field configuration varies with the beat period of the two spiral patterns.
The configuration of Fig.~\ref{fig:I} is chosen arbitrarily, 
but the discussion is valid for all times in the saturated regime.
The field is axisymmetric near the centre but magnetic arms (blue)
emerge near the corotation radii of the patterns (dotted circles), 
interlaced with the gaseous (enhanced $U_z$) arms (black contours).
This is more evident in the lower panel of Fig.~\ref{fig:I}, 
where the degree of non-axisymmetry $\delta\equiv[B_\phi-B_\phi^{(0)}(r)]/B_\phi^{(0)}(r)$ is shown, 
with the superscript denoting the azimuthal wave number, $m=0$ for the axially symmetric part of the field
(we note that $B_\phi$ generally dominates over $B_r$ for the models considered here).
The magnetic arms, here in white--yellow colours, are rather strong,
$|\delta|\leq0.6$, but localised within a few kpc of the corotation region. 
It is worth mentioning that the large-scale magnetic field is found to be concentrated in the interarm regions
even for a simpler spiral model that consists of a single, rigidly rotating pattern,
but in that case magnetic arms are found to be weaker and less radially extended.

As shown in Fig.~\ref{fig:D}, Model~D (shown for $\tau\rightarrow0$) also produces 
fairly strong ($\delta=0.3$--$0.4$) magnetic arms interlaced with the gaseous ones, 
but extended in radius with pitch angles similar to those of the gaseous spiral. 
Such a situation is quite close to what is seen in NGC~6946.
The degree of deviation from axial symmetry varies with time as the spiral winds up.
Fig.~\ref{fig:delta_time} shows that the maximum value of $\delta$ first increases with time 
for models with $\tau\rightarrow0$ (solid) and $\tau=l/u$ (dashed),
and peaks a few hundred $\Myr$ after the onset of the spiral.
Evidently, stronger magnetic arms are produced when $\tau$ is finite \citep{Chamandy14};
even for this case they are concentrated in between the gaseous arms.
The location of the maximum of $\delta$ moves outward with time, 
from $r=6.0\kpc$ ($5.8\kpc$) at $t-t\f=100\Myr$ to $r=8.2\kpc$ ($7.8\kpc$) at $500\Myr$ for $\tau\rightarrow0$ ($\tau=l/u$).
This outward propagation is explained by the increase with radius of the local response time to the spiral perturbation \citep{Chamandy+13a}.

Predictably, smaller $\epsilon$ (weaker non-axisymmetric forcing) produces weaker magnetic arms,
with $\delta\ma$ reduced in proportion to $\epsilon$ in Model~D.
Changing $\kappa$ has little effect on $\delta$ but reduces $B$ by more than a factor
of two as $\kappa$ is reduced from 1 to 0, i.e., when the flux of $\alpha\magn$ through
the disc surface is reduced to that due to advection alone.
The effect of changing $U\f$ is more complicated (see Section~\ref{sec:outflow}).
For $\kappa=1$ and $\tau=l/u$, $\delta$ decreases substantially (almost by 40 per cent) as $U\f$ is 
reduced (from 3 to $2\kms$), while $B$ increases slightly ($\simeq10$ per cent).
For the mechanism to be viable, the outflow must be strong enough to affect the dynamo in the 
gaseous arms, but not strong enough to suppress it globally.
Our results are not very sensitive to the degree of flaring.
For an unflared disc with $h=0.5\kpc$, 
$h$ is larger (smaller) than the flared disc inside (outside) $r=8\kpc$;
this leads to a slight reduction in $\delta$ for $r<8\kpc$ and a slight increase for $r>8\kpc$.
This finding is not surprising because larger scale height translates to a more supercritical dynamo number,
and thus results in weaker magnetic arms, as explained in Section~\ref{sec:outflow}.

\section{Conclusions}
\label{sec:conclusion}
We have shown that magnetic arms situated in between the gaseous spiral arms can be generated when 
the mean outflow speed $U_z$ is stronger in the gaseous arms, independently of the spiral model used.
This assumption of spiral modulation of the outflow speed is supported, to some extent at least, by observation and theory,
and its impact on the dynamo is simple and direct.
In particular, if the gaseous arms wind up as transient density waves,
an interlaced pattern of magnetic and gaseous arms can persist in a wide radial range, as observed in some galaxies.
The fact that at least some observations can be better explained using such models of spiral structure and evolution lends support to those models.

The mechanism proposed is most effective when the dynamo is close to critical.
If stronger magnetic fields enhance the formation rate of massive stars \citep{Mestel99,Dobbs+13},
leading to stronger outflows,
the dynamo could be self-regulated to remain near critical.
In any case, our tentative prediction is that galaxies with higher star formation rates, 
and hence stronger outflows, are more likely to possess magnetic arms in between the gaseous ones. 
Indeed, the galaxies in which interarm magnetic arms have been identified are found to be gas-rich \citep{Beck+Wielebinski13}.
We intend to extend our models to include the three-dimensional structures of the disc and outflow
and observationally constrained parameter values for specific galaxies.

\section*{Acknowledgments}
We are grateful to R.-J.~Dettmar for a useful discussion. A.S.\ gratefully acknowledges financial
support of IUCAA and STFC (grant ST/L005549/1).

\footnotesize{
\noindent
\bibliographystyle{mn2e}
\bibliography{refs}
}

\label{lastpage}
\end{document}